# Decentralized and Collaborative Subspace Pursuit: A Communication-Efficient Algorithm for Joint Sparsity Pattern Recovery with Sensor Networks

Gang Li, *Senior Member, IEEE*, Thakshila Wimalajeewa, *Member, IEEE*, and Pramod K. Varshney, *Fellow, IEEE*

**Abstract:** In this paper, we consider the problem of joint sparsity pattern recovery in a distributed sensor network. The sparse multiple measurement vector signals (MMVs) observed by all the nodes are assumed to have a common (but unknown) sparsity pattern. To accurately recover the common sparsity pattern in a decentralized manner with a low communication overhead of the network, we develop an algorithm named decentralized and collaborative subspace pursuit (DCSP). In DCSP, each node is required to perform three kinds of operations per iteration: 1) estimate the local sparsity pattern by finding the subspace that its measurement vector most probably lies in; 2) share its local sparsity pattern estimate with one-hop neighboring nodes; and 3) update the final sparsity pattern estimate by majority vote based fusion of all the local sparsity pattern estimates obtained in its neighborhood. The convergence of DCSP is proved and its communication overhead is quantitatively analyzed. We also propose another decentralized algorithm named generalized DCSP (GDCSP) by allowing more information exchange among neighboring nodes to further improve the accuracy of sparsity pattern recovery at the cost of increased communication overhead. Experimental results show that, 1) compared with existing decentralized algorithms, DCSP provides much better accuracy of sparsity pattern recovery at a comparable communication cost; and 2) the accuracy of GDCSP is very close to that of centralized processing.

---

G. Li is with the Department of Electronic Engineering, Tsinghua University, Beijing 100084, China. Email: gangli@tsinghua.edu.cn .

T. Wimalajeewa and P. K. Varshney are with the Department of Electrical Engineering and Computer Science, Syracuse University, Syracuse, NY 13244, USA. Email: {twwewelw, varshney}@syr.edu

The work of G. Li was supported in part by the National Natural Science Foundation of China under Grants 61422110 and 41271011, and in part by the Tsinghua University Initiative Scientific Research Program. The work of T. Wimalajeewa and P. K. Varshney was supported in part by National Science Foundation Award No. 1307775.

## I. INTRODUCTION

Compressive sensing (CS) refers to the idea that a sparse signal can be accurately recovered from a small number of measurements [1]-[3]. It has been shown that CS is potentially useful in a wide range of applications including medical imaging [4][5], radar imaging [6]-[8], and source localization [9]-[11]. In particular, CS provides a new approach for data reduction in sensor network applications without compromising performance [12]-[15].

Assume that the sparse multiple measurement vector signals (MMVs) are observed by a sensor network consisting of $Q$ spatially distributed nodes. The measurements collected at the $q$-th node are given by

$$\mathbf{y}_q = \mathbf{A}_q \mathbf{x}_q \tag{1}$$

where $\mathbf{y}_q$ is an $M$×1 measurement vector, $\mathbf{A}_q$ is an $M$×$N$ dictionary matrix assumed to be fixed at a given node, $\mathbf{x}_q$ is an $N$×1 vector which has $K$ nonzero entries, $q$=1, 2, $\cdots$, $Q$. Assume that all {$\mathbf{x}_q$, $q$=1, 2, $\cdots$, $Q$} have the same sparsity pattern, i.e., the support set is defined as $T$ ={$i$: $\mathbf{x}_q(i) \neq 0$, $i$=1, 2, $\cdots$, $N$ } with cardinality $|T|$=$K$. Here we consider the case that {$\mathbf{x}_q(T)$, $q$=1, 2, $\cdots$, $Q$} are random vectors and they are statistically independent of each other. This scenario is applicable when multiple sensor nodes collect independent snapshots [30], and when multiple heterogeneous sensors, e.g., radar and camera, or radars with different frequencies and angles of incidence, observe the same group of targets [16][17][31][32]. Our goal is to estimate the support set $T$ using {$\mathbf{y}_q$, $\mathbf{A}_q$, $q$=1, 2, $\cdots$, $Q$} in the case when $N$>$M$≥2$K$.

In a centralized processing framework, it is required for each node to transmit its local measurement data and the information regarding the local dictionary matrix to a central processor. Typical centralized algorithms include the simultaneous orthogonal matching pursuit (SOMP) [18][19] and the simultaneous subspace pursuit (SSP) algorithms [20]. Due to practical constraints on communication bandwidth and computational capacity, recovering the common sparsity pattern in a decentralized manner is more efficient. In decentralized processing, some preliminary processing is carried out at each node based on its local measurement data, and then the performance is improved by the collaboration among multiple nodes. In decentralized convex optimization algorithms [21]-[25], the neighboring nodes are required to share their $N$-length local solutions with each other per iteration to update the parameters of the local optimization problems. Among them, the distributed alternating direction method of multipliers (D-ADMM) approach [23] requires less communication overhead than others. The bandwidth usage in [21]-[25] is considerably high, because the $N$-length local solution to be transmitted from each node is probably not sparse before the iterative process converges. In [26], the distributed iterative hard thresholding (DIHT) algorithm was proposed for tree networks, in which collaboration among the nodes is carried out via the broadcast of $K$-length messages from the parent nodes and the convergecast of $N$-length messages from the children nodes. Thanks to the reduced number of transmissions of $N$-length messages, DIHT has less communication overhead than the methods in [21]-[25]. In the scenario where the communication capacity of the network is extremely limited so that the transmission of $N$-length messages is difficult to realize, the algorithms in [21]-[26] may fail to work. Distributed orthogonal matching pursuit (DiOMP) and distributed subspace pursuit (DiSP) [27] work well with a network that has very limited communication bandwidth, because they only require the transmission of $K$-length locally selected indices among neighboring nodes. Two versions of the decentralized and collaborative orthogonal matching pursuit (DCOMP) algorithm were proposed in [28]. DCOMP 1 only requires the transmission of one index among neighboring nodes at a given iteration, so its communication overhead is very low at the cost of accuracy loss. In DCOMP 2, which requires each node to share $N$-length correlation coefficients with its neighbors and one index with other remote nodes in the network at a given iteration, respectively, the accuracy of sparsity pattern estimation is significantly improved at the cost of increased communication overhead.

In this paper, by embedding fusion among nodes into the iterative process of the standard subspace pursuit algorithm [33], we develop a decentralized algorithm named decentralized subspace pursuit (DCSP) for joint sparsity pattern recovery under communication constrains. The current work is based on our preliminary work [29]. Here we significantly extend the work in [29] by providing theoretical analysis on the performance of DCSP and results based on more detailed experiments. The proposed DCSP algorithm operates in an iterative fashion. Each iteration of DCSP contains three operations. First, each node selects $K$ columns from the local dictionary matrix to approximate the subspace that the local measurement vector most probably lies in. Then each node sends the indices of $K$ selected columns to its one-hop neighbors. Finally, each node updates the support set estimate by majority vote based fusion of all the indices selected in its neighborhood. DCSP only requires the exchange of $K$-length indices among neighboring nodes, which is useful in networks with extremely limited bandwidth. The convergence of DCSP is proved and the communication overhead of DCSP is quantitatively analyzed. Experimental results show that, compared to existing decentralized greedy algorithms such as DiOMP and DiSP, DCSP provides much better accuracy of sparsity pattern recovery at a comparable communication cost. We further propose a generalized DCSP (GDSCP) algorithm, in which O($N$)-length messages are allowed to be transmitted in a small neighborhood surrounding each node. GDCSP is superior to DCSP in terms of the accuracy of sparsity pattern recovery, at the cost of increased communication overhead. As demonstrated by numerical experiments, compared to the DCOMP algorithm we developed before in [28], both DCSP and GDCSP further reduce the communication overhead over the network without sacrificing the accuracy of sparsity pattern recovery.

The rest of this paper is organized as follows. In Section II, we describe the implementation of DCSP for joint sparsity pattern recovery in a decentralized sensor network. In Section III, we discuss the convergence of DCSP and its communication overhead. The GDCSP algorithm is proposed in Section IV. Simulation results are provided in Section V and concluding remarks are given in Section VI.

*Notations*: To simplify the presentation, we define the following notations used in this paper. $\mathrm{ls}(\mathbf{y},\mathbf{A}) = [\mathbf{A}^H\mathbf{A}]^{-1}\mathbf{A}^H\mathbf{y}$ provides the projection coefficients of a vector $\mathbf{y}$ onto the column space of a matrix $\mathbf{A}$, where conjugate transpose is denoted by $(\cdot)^H$; $\mathrm{resid}(\mathbf{y},\mathbf{A}) = \mathbf{y} - \mathbf{A}[\mathbf{A}^H\mathbf{A}]^{-1}\mathbf{A}^H\mathbf{y}$ outputs the projection residual vector; max_ind($\mathbf{y}$, $K$)={$K$ indices corresponding to the largest magnitude entries in the vector $\mathbf{y}$}; max_occ($T$, $K$) ={$K$ elements that have the highest frequency of occurrence in the set $T$ }; $\mathbf{A}(T)$ denotes a sub-matrix composed of the columns of $\mathbf{A}$ indexed by the set $T$; $\mathbf{y}(T)$ denotes a sub-vector composed of the entries of $\mathbf{y}$ indexed by the set $T$; $|\mathbf{y}|^n$ computes the $n$-th power of the absolute value of the vector $\mathbf{y}$ element-by-element; $|T|$ denotes the cardinality of the set $T$; the

set-subtraction $T - S$ outputs a set composed of those elements belonging to $T$ but not $S$.

## II. ALGORITHM DESCRIPTION

Throughout this paper, we consider an undirected connected graph $\Upsilon = (G, D)$ with node set $G = \{1,2,\ldots,Q\}$ and edge set $D$, where $(i, j) \in D$ means that one-hop communication between nodes $i$ and $j$ is available. Define $G_q = \{k \mid (k,q) \in D\} \bigcup \{q\}$ to be the set containing the indices of the $q$-th node itself and its one-hop neighboring nodes. The operations of the DCSP algorithm at the $q$-th node are summarized in Algorithm 1. In the initialization phase, the local estimate of the support set is obtained by finding $K$ maximum correlation coefficients between the local measurement vector and the columns of the local dictionary matrix. After exchanging $K$ selected indices with the neighboring nodes, the $q$-th node collects $K|G_q|$ indices and records them in the set $\Gamma_q^0$. Then at the $q$-th node, the support set estimate is updated by finding $K$ elements in $\Gamma_q^0$ that have the highest frequency of occurrence. After initialization, in Step 4 of DCSP, each node first enlarges the set of index candidates and then selects $K$ indices corresponding to the largest projection coefficients. This finds a $K$-dimensional subspace that the local measurement vector most likely lies in. By exchanging $K$ selected indices with the neighboring nodes, the set $\Gamma_q^l$ at the $q$-th node consists of $K|G_q|$ indices. Some elements of $\Gamma_q^l$ may occur more than once, which means that some indices are selected by multiple nodes. In Step 5 of DCSP, index fusion is carried out by majority voting, i.e., $K$ elements in $\Gamma_q^l$ that have the highest frequency of occurrence are selected as the new estimate of the support set. In Step 6 of DCSP, the residual is updated by using the new estimate of the support set. In Step 7 of DCSP, the stopping criterion is checked. When the residual error is not decreased at the current iteration, the support set estimate and the residual vector obtained at the previous iteration are deemed reliable, and the flag parameter $f(q)$ is set to 1. The iterative process at the $q$-th node is terminated if no updating of support set estimate occurs in its surrounding neighborhood.

---

**Algorithm 1 The DCSP algorithm at the $q$-th node**

Input: $K$, $\mathbf{y}_q$ , $\mathbf{A}_q$.

Initialization:

1) Transmit $\Omega_q^0$ =max_ind( $|\mathbf{A}_q^H \mathbf{y}_q|$ , $K$) to and receive $\Omega_j^0$ from the $j$-th node for all $j \in G_q \setminus \{q\}$.

2) Let $\Gamma_q^0 = \left\{\Omega_j^0, j \in G_q\right\}$ and $T_q^0$ =max_occ( $\Gamma_q^0$ , $K$) .

3) Calculate the local residual $\mathbf{r}_q^0$ = resid($\mathbf{y}_q$, $\mathbf{A}_q$( $T_q^0$ )).

Iteration: at the $l$-th iteration ($l \geqslant 1$)

4) Set $f(q)$ = 0. Let $\Omega_q^l$ =max_ind( $\left|\text{ls}(\mathbf{y}_q, \mathbf{A}_q(\bar{\Omega}_q^l))\right|$ , $K$), where $\bar{\Omega}_q^l = T_q^{l-1} \cup$ max_ind ( $|\mathbf{A}_q^H \mathbf{r}_q^{l-1}|$ , $K$) . Transmit $\Omega_q^l$ to and receive $\Omega_j^l$ from the $j$-th node for all $j \in G_q \setminus \{q\}$.

5) Let $\Gamma_q^l = \left\{\Omega_j^l, j \in G_q\right\}$ and $T_q^l$ =max_occ( $\Gamma_q^l$ , $K$).

6) Updates the local residual $\mathbf{r}_q^l$ = resid($\mathbf{y}_q$, $\mathbf{A}_q$( $T_q^l$ )).

7) If $\left\|\mathbf{r}_q^l\right\|_2 \geq \left\|\mathbf{r}_q^{l-1}\right\|_2$ , let $T_q^l = T_q^{l-1}$, $\mathbf{r}_q^l = \mathbf{r}_q^{l-1}$, and $f(q)$ =1. Send $f(q)$ to and receive $f(j)$ from the $j$-th node for all $j \in G_q \setminus \{q\}$. If $f(j)$ = 1 for all $j \in G_q$, stop the iteration; otherwise, let $l$=$l$+1 and return to Step 4.

Output: The estimated support set $\hat{T}_q = T_q^l$ .

---

## III. PERFORMANCE ANALYSIS OF DCSP

In this section, we prove the convergence of DCSP and analyze its communication overhead. We first provide a review of the concept of the restricted isometry property (RIP).

***Definition 1*** RIP [1][2]: A $M \times N$ matrix $\mathbf{A}$ satisfies the RIP with parameters ($K$, $\delta$ ) for $K \leqslant M$ and $0 \leqslant \delta < 1$, if

$$(1-\delta)\left\|\mathbf{x}\right\|_2^2 \leq \left\|\mathbf{A}\mathbf{x}\right\|_2^2 \leq (1+\delta)\left\|\mathbf{x}\right\|_2^2 \tag{2}$$

holds for any sparse signals $\mathbf{x}$ whose $l_0$ pseudo-norm is not larger than $K$, where $\left\|\cdot\right\|_2$ denotes the $l_2$ norm.

### *A. Convergence of DCSP*

Now we investigate the convergence of DCSP in the measurement-noise-free case. At the $l$-th iteration, define a $N \times 1$ binary vector $\boldsymbol{\beta}_q^l$ by

$$\boldsymbol{\beta}_q^l(i)=\begin{cases}1, & \text{if } i\in\Omega_q^l\\ 0, & \text{otherwise}\end{cases}, \tag{3}$$

for $i$=1,2, ⋯, $N$. That is to say, the entries 1's in $\boldsymbol{\beta}_q^l$ are indexed by the set $\Omega_q^l$. Let

$$\boldsymbol{\alpha}_q^l=\sum_{j\in G_q}\boldsymbol{\beta}_j^l . \tag{4}$$

In Steps 2 and 5 of DCSP, the fusion based on majority vote is performed by finding $K$ elements that have the highest frequency of occurrence in $\Gamma_q^l$. This is equivalent to saying that $T_q^l$ is composed of $K$ indices corresponding to the largest coefficients in $\boldsymbol{\alpha}_q^l$. The proof of convergence of DCSP is carried out by the following two steps: evaluating the reliabilities of $T_q^0$ at the initialization phase and $T_q^l$ at the $l$-th iteration.

*1) The reliability of $T_q^0$ at the initialization phase of DCSP*

We first investigate the effect of collaboration among the nodes on the accuracy of support set estimation in the initialization phase of DCSP. Since $T_q^0$ is composed of $K$ indices corresponding to the largest coefficients in $\boldsymbol{\alpha}_q^0$, the probability $\Pr\left(\boldsymbol{\alpha}_q^0(n)>\boldsymbol{\alpha}_q^0(m)\right)$ for $n\in T$ and $m\in T_c$ represents the reliability of $T_q^0$.

***Proposition 1*** When the dictionary matrix $\mathbf{A}_q$ satisfies the RIP with the constant $\delta_{3K}\le 0.206$, $N\ge K^2$, and the number of one-hop neighbors of the $q$-th node satisfies

$$\left|G_q\right|\ge\frac{9.21}{\left(\dfrac{1}{K}-\dfrac{K-1}{N-K}\right)^2}, q=1,2,\cdots,Q , \tag{5}$$

we have $\Pr\left(\boldsymbol{\alpha}_q^0(n)>\boldsymbol{\alpha}_q^0(m)\right)\ge 0.99$ for $n\in T$ and $m\in T_c$, where the superscript 0 denotes the initialization phase of DCSP.

***Proof*:** At Step 1 of DCSP, $K$ indices corresponding to the largest correlation coefficients between $\mathbf{y}_q$ and the columns of $\mathbf{A}_q$ are recorded in the set $\Omega_q^0$. After the local selection of indices at Step 1 of DCSP, we have

$$\left\|\mathbf{x}_q(T-\Omega_q^0)\right\|_2\le\frac{\sqrt{8\delta_{2K}-8\delta_{2K}^2}}{1+\delta_{2K}}\left\|\mathbf{x}_q\right\|_2\le\frac{\sqrt{8\delta_{3K}-8\delta_{3K}^2}}{1+\delta_{3K}}\left\|\mathbf{x}_q\right\|_2\le 0.952\left\|\mathbf{x}_q\right\|_2 . \tag{6}$$

Here, the first inequality holds when $\delta_{2K}\le 0.5$ (see Appendix C in [33]), the second and the third inequalities hold since $\delta_{2K}\le\delta_{3K}$ [33]. This implies that, some elements in $\Omega_q^0$ belong to $T$, i.e., $T\cap\Omega_q^0\ne\varnothing$, for $q$=1, 2, ⋯, $Q$. Let $K_q^0 = |T \cap \Omega_q^0|$, i.e., $K_q^0$ indices in $\Omega_q^0$ belong to *T*. It is obvious that $1 \le K_q^0 \le K$. Accordingly, the other $K - K_q^0$ indices in $\Omega_q^0$ belong to $T_c = \{1, 2, \cdots, N\} \setminus T$. Note that $\Omega_q^0$ = max_ind($|\mathbf{A}_q^H \mathbf{y}_q|$, *K*) = max_ind($|\mathbf{A}_q^H \mathbf{A}_q(T)\mathbf{x}_q(T)|$, *K*). Due to the randomness of the magnitudes of $\mathbf{x}_q(T)$, there is no *a priori* knowledge regarding some indices in *T* that have higher priority to be selected into $\Omega_q^0$ than others. According to Laplace's principle of insufficient reason [40], the indices in *T* have an equal opportunity to be selected into $\Omega_q^0$, i.e., the elements of $|T \cap \Omega_q^0|$ are uniformly distributed within *T*. Similarly, the elements of $|T_c \cap \Omega_q^0|$ are also uniformly distributed within $T_c$. Therefore, $\boldsymbol{\beta}_q^0(n)$ with $n \in T$ and $\boldsymbol{\beta}_q^0(m)$ with $m \in T_c$ follow Bernoulli distributions:

$$\begin{aligned} \Pr\left(\boldsymbol{\beta}_q^0(n)=1\right) &= \frac{C_{K-1}^{K_q^0-1}\cdot C_{N-K}^{K-K_q^0}}{C_K^{K_q^0}\cdot C_{N-K}^{K-K_q^0}} = \frac{K_q^0}{K} \ge \frac{\tilde{K}_q^0}{K} \\ \Pr\left(\boldsymbol{\beta}_q^0(n)=0\right) &= 1-\Pr\left(\boldsymbol{\beta}_q^0(n)=1\right) = 1-\frac{K_q^0}{K} \le 1-\frac{\tilde{K}_q^0}{K} \end{aligned}, \quad n \in T, \tag{7}$$

and

$$\begin{aligned} \Pr\left(\boldsymbol{\beta}_q^0(m)=1\right) &= \frac{C_K^{K_q^0}\cdot C_{N-K-1}^{K-K_q^0-1}}{C_K^{K_q^0}\cdot C_{N-K}^{K-K_q^0}} = \frac{K-K_q^0}{N-K} \le \frac{K-\tilde{K}_q^0}{N-K} \\ \Pr\left(\boldsymbol{\beta}_q^0(m)=0\right) &= 1-\Pr\left(\boldsymbol{\beta}_q^0(m)=1\right) = 1-\frac{K-K_q^0}{N-K} \ge 1-\frac{K-\tilde{K}_q^0}{N-K} \end{aligned}, \quad m \in T_c, \tag{8}$$

where

$$\tilde{K}_q^0 \triangleq \min_{j \in G_q}\{K_j^0\} \tag{9}$$

and $C_r^j = \dfrac{r!}{j!(r-j)!}$. The definition of $\tilde{K}_q^0$ implies that each node in the neighborhood of the *q*-th node has correctly selected at least $\tilde{K}_q^0$ indices after Step 1 of DCSP. This is also equivalent to saying that at most $K - \tilde{K}_q^0$ indices in $\Omega_j^0$ are incorrect for $j \in G_q$, since all the $\Omega_j^0$ for $j \in G_q$ have the same cardinality *K*.

From (4) we have

$$\Pr\left(\boldsymbol{\alpha}_q^0(n) > \boldsymbol{\alpha}_q^0(m)\right) = \Pr\left(\sum_{j \in G_q}\left(\boldsymbol{\beta}_j^0(n) - \boldsymbol{\beta}_j^0(m)\right) > 0\right). \tag{10}$$

Let $s^0 = \boldsymbol{\alpha}_q^0(n) - \boldsymbol{\alpha}_q^0(m) = \sum_{j \in G_q}\left(\boldsymbol{\beta}_j^0(n) - \boldsymbol{\beta}_j^0(m)\right)$, then

$$E(s^0) = \sum_{j \in G_q}\left(\frac{K_j^0}{K} - \frac{K-K_j^0}{N-K}\right) = \frac{1}{K(N-K)}\sum_{j \in G_q}(NK_j^0 - K^2). \tag{11}$$

where $E(\cdot)$ denotes the expectation operator. Note that $\Omega_q^0$=max_ind($|\mathbf{A}_q^H\mathbf{y}_q|$, $K$)=max_ind($|\mathbf{A}_q^H\mathbf{A}_q(T)\mathbf{x}_q(T)|$, $K$). We can conclude that {$\Omega_q^0$, $q$=1,2, ⋯, $Q$} and, therefore, {$\boldsymbol{\beta}_q^0$, $q$=1,2, ⋯, $Q$} are statistically independent due to the independence among the vectors {$\mathbf{x}_q(T)$, $q$=1,2, ⋯, $Q$}. Since $\left(\boldsymbol{\beta}_j^0(n)-\boldsymbol{\beta}_j^0(m)\right)\in\{-1,0,1\}$, the Hoeffding's tail inequality [38][39] tells us that

$$\Pr\left(s^0-E(s^0)\le-\omega\right)\le e^{-\frac{\omega^2}{2|G_q|}} \tag{12}$$

for $\omega>0$. By choosing $\omega=E(s^0)$ and substituting (11) into (12), we have

$$\begin{aligned}\Pr\left(s^0\le 0\right)&\le\exp\left(-\frac{\left(\sum_{j\in G_q}(NK_j^0-K^2)\right)^2}{2|G_q|K^2(N-K)^2}\right)\\&\le\exp\left(-\frac{|G_q|(N\tilde{K}_q^0-K^2)^2}{2K^2(N-K)^2}\right)\quad,\\&=\exp\left(-\frac{|G_q|\left(p_{1,q}^0-p_{2,q}^0\right)^2}{2}\right)\end{aligned} \tag{13}$$

where

$$\begin{aligned}p_{1,q}^0&\triangleq\frac{\tilde{K}_q^0}{K}\\p_{2,q}^0&\triangleq\frac{K-\tilde{K}_q^0}{N-K}\end{aligned}. \tag{14}$$

The second inequality in (13) holds because of (9) and the assumption $N\ge K^2$. Also under the assumption $N\ge K^2$, elementary calculations yield

$$p_{1,q}^0=\frac{\tilde{K}_q^0}{K}\ge\tilde{K}_q^0\cdot\frac{K-1}{N-K}\ge\frac{K-\tilde{K}_q^0}{N-K}=p_{2,q}^0\,. \tag{15}$$

From (13), we have

$$\Pr\left(\boldsymbol{\alpha}_q^0(n)>\boldsymbol{\alpha}_q^0(m)\right)=\Pr\left(s^0>0\right)\ge1-e^{-\frac{|G_q|\left(p_{1,q}^0-p_{2,q}^0\right)^2}{2}}\,. \tag{16}$$

Based on (16), the sufficient condition for $\Pr\left(\boldsymbol{\alpha}_q^0(n)>\boldsymbol{\alpha}_q^0(m)\right)\ge0.99$ can be given by

$$|G_q|\ge\frac{9.21}{\left(p_{1,q}^0-p_{2,q}^0\right)^2}=\frac{9.21}{\left(\frac{\tilde{K}_q^0}{K}-\frac{K-\tilde{K}_q^0}{N-K}\right)^2}\,. \tag{17}$$

The value of $\tilde{K}_q^0$ may vary for different kinds of sparse signals. Substituting $\tilde{K}_q^0=1$ into (17) yields (5), which is the sufficient condition that $\Pr\left(\boldsymbol{\alpha}_q^0(n) > \boldsymbol{\alpha}_q^0(m)\right) \geq 0.99$ holds for any sparse signals. This completes the proof of Proposition 1. ■

Since $T_q^0$ is composed of $K$ indices corresponding to the largest coefficients in $\boldsymbol{\alpha}_q^0$, from Proposition 1 it is clear that, when the number of neighboring nodes satisfies (5), every element in $T_q^0$ is correct with high probability. Thus, it can be concluded that

$$\left\|\mathbf{x}_q(T - T_q^0)\right\|_2 \leq \left\|\mathbf{x}_q(T - \Omega_q^0)\right\|_2 \tag{18}$$

holds with high probability when the value of $|G_q|$ is large enough, i.e., the collaboration among nodes improves the accuracy of support set estimation.

Besides the above theoretical analysis, in what follows we will further determine $\Pr\left(\boldsymbol{\alpha}_q^0(n) > \boldsymbol{\alpha}_q^0(m)\right)$ versus the value of $|G_q|$ via numerical computations. For $n \in T$ and $m \in T_c$, the covariance between $\boldsymbol{\beta}_q^l(n)$ and $\boldsymbol{\beta}_q^l(m)$ is given by

$$\begin{aligned} \operatorname{cov}\left(\boldsymbol{\beta}_q^0(n), \boldsymbol{\beta}_q^0(m)\right) &= E\left(\boldsymbol{\beta}_q^0(n) \cdot \boldsymbol{\beta}_q^0(m)\right) - E\left(\boldsymbol{\beta}_q^0(n)\right) \cdot E\left(\boldsymbol{\beta}_q^0(m)\right) \\ &= \frac{C_{K-1}^{K_q^0-1} \cdot C_{N-K-1}^{K-K_q^0-1}}{C_K^{K_q^0} \cdot C_{N-K}^{K-K_q^0}} - \frac{K_q^0}{K} \cdot \frac{K - K_q^0}{N-K} \\ &= 0 \end{aligned} \tag{19}$$

As proved in [35], two Bernoulli random variables are statistically independent if and only if they are uncorrelated. Thus, (19) implies that $\boldsymbol{\beta}_q^0(n)$ and $\boldsymbol{\beta}_q^0(m)$ are statistically independent of each other for $n \in T$ and $m \in T_c$. Accordingly, it follows that $\boldsymbol{\alpha}_q^0(n)$ and $\boldsymbol{\alpha}_q^0(m)$ are statistically independent for $n \in T$ and $m \in T_c$. From (13) and (16), we can see that the probability $\Pr\left(\boldsymbol{\alpha}_q^0(n) > \boldsymbol{\alpha}_q^0(m)\right)$ reaches its lower bound when $K_j^0 = \tilde{K}_q^0$ for all $j \in G_q$. Thus, we have

$$\begin{aligned} \Pr\left(\boldsymbol{\alpha}_q^0(n) > \boldsymbol{\alpha}_q^0(m)\right) &= \sum_{i=1}^{|G_q|} \sum_{j=0}^{i-1} \Pr\left(\boldsymbol{\alpha}_q^0(n) = i, \boldsymbol{\alpha}_q^0(m) = j\right) \\ &= \sum_{i=1}^{|G_q|} \sum_{j=0}^{i-1} \Pr\left(\boldsymbol{\alpha}_q^0(n) = i\right) \cdot \Pr\left(\boldsymbol{\alpha}_q^0(m) = j\right) \\ &\overset{K_j^0 = \tilde{K}_q^0}{\geq} \sum_{i=1}^{|G_q|} C_{|G_q|}^i (p_{1,q}^0)^i (1 - p_{1,q}^0)^{|G_q|-i} \left( \sum_{j=0}^{i-1} C_{|G_q|}^j (p_{2,q}^0)^j (1 - p_{2,q}^0)^{|G_q|-j} \right) \end{aligned} \tag{20}$$

According to (20), given different values of $\tilde{K}_q^0 / K$ and $N / K$, in Fig. 1 we plot the lower bound of $\Pr\left(\boldsymbol{\alpha}_q^0(n) > \boldsymbol{\alpha}_q^0(m)\right)$ for $n \in T$ and $m \in T_c$ versus the number of one-hop neighboring nodes of the $q$-th node. We can see that, to ensure that $\Pr\left(\boldsymbol{\alpha}_q^0(n) > \boldsymbol{\alpha}_q^0(m)\right) \geq 0.99$ for $n \in T$ and $m \in T_c$, 13, 17, 47, and more than 50 neighboring nodes are required when the parameter pair ( $\tilde{K}_q^0 / K$ , $N / K$ ) is equal to (0.4, 20), (0.4, 10), (0.2, 20) and (0.2, 10), respectively. It is clear that the required value of $|G_q|$ decreases with the increasing values of $\tilde{K}_q^0 / K$ and $N / K$. This conclusion is straightforward from the following considerations: 1) the more correct indices there are in the local estimate of the support set, the fewer the nodes required to achieve a desired accuracy of fusion; and 2) the sparser the signal is, the smaller the probability that different nodes select the same wrong indices from $T_c$. Fig. 1 demonstrates that, when $\tilde{K}_q^0 > 1$, the number of neighbors of each node required to yield high probability $\Pr\left(\boldsymbol{\alpha}_q^0(n) > \boldsymbol{\alpha}_q^0(m)\right)$ for $n \in T$ and $m \in T_c$ can be much smaller than that given by (5).

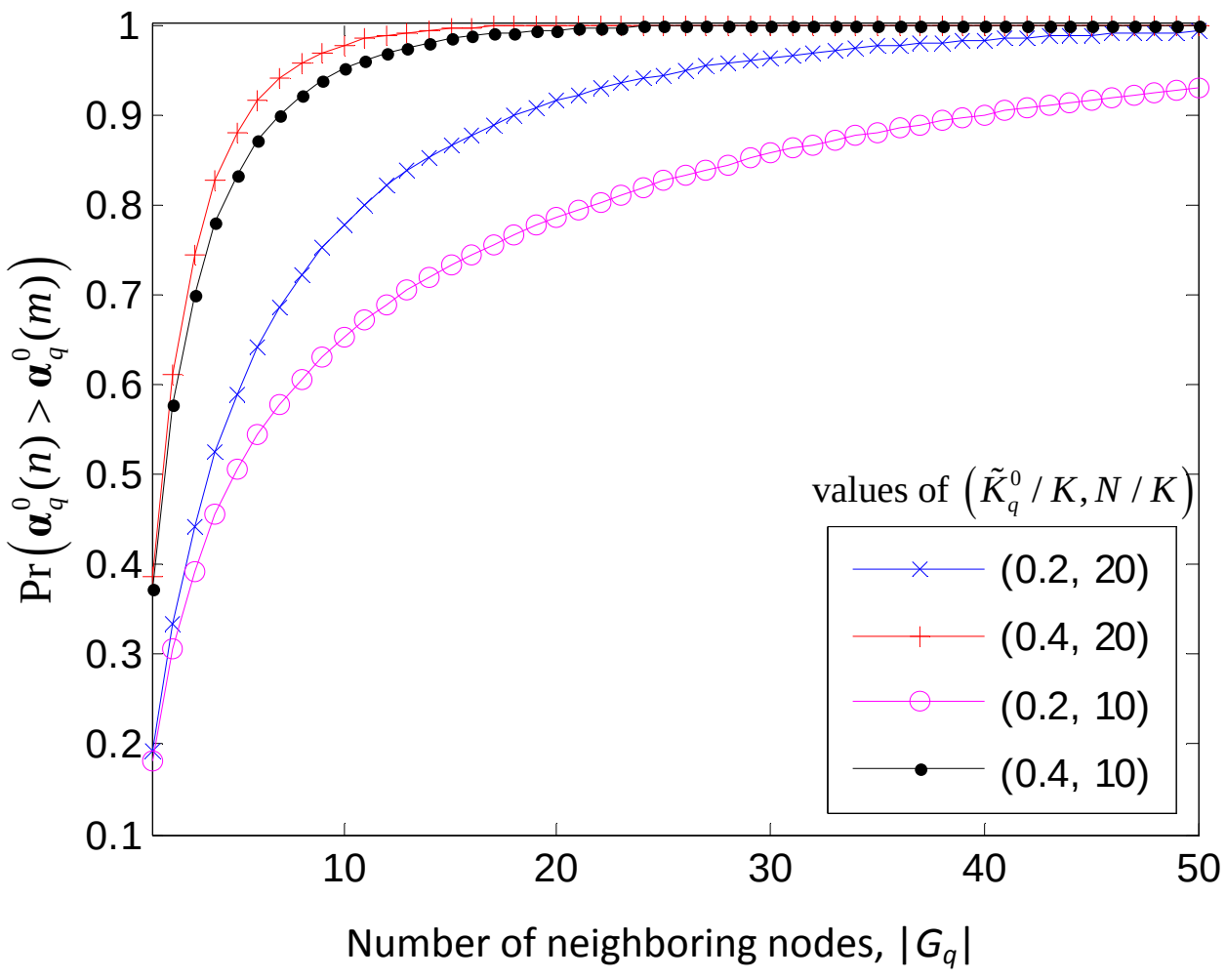

Fig. 1 The probability $\Pr\left(\boldsymbol{\alpha}_q^0(n) > \boldsymbol{\alpha}_q^0(m)\right)$ for $n \in T$ and $m \in T_c$ versus the value of $|G_q|$.

*2) The reliability of $T_q^l$ at the l-th iteration for $l \geq 1$ of DCSP*

Next, we discuss the reliability of $T_q^l$ at the $l$-th iteration of DCSP with $l \geqslant 1$. In Step 4 of DCSP, the set of index candidates is enlarged and a new group of $K$ indices corresponding to the largest projection coefficients is selected at each node. By doing so, it is possible to remove some wrong indices that were considered reliable during past iterations and to add some new indices into the support set estimate, i.e., the reliability of the support set estimate is reevaluated based on the local measurement data. This step is the same as that in the standard SP algorithm, so according to Theorem 3 and Theorem 4 in [33] we have

$$\left\|\mathbf{x}_q(T-\Omega_q^l)\right\|_2 \le \frac{2\delta_{3K}(1+\delta_{3K})}{(1-\delta_{3K})^3}\left\|\mathbf{x}_q(T-T_q^{l-1})\right\|_2 < \left\|\mathbf{x}_q(T-T_q^{l-1})\right\|_2 . \tag{21}$$

where the second inequality holds since the RIP constant $\delta_{3K} \le 0.206$ . From (18) and (21) it follows that

$$\left\|\mathbf{x}_q(T-\Omega_q^l)\right\|_2 < \left\|\mathbf{x}_q(T-\Omega_q^{l-1})\right\|_2 \tag{22}$$

holds with high probability.

Assume that, at the *l*-th iteration, the *q*-th node has correctly selected $K_q^l$ indices that belong to the true support set after Step 4 of DCSP, i.e., $K_q^l = |T \cap \Omega_q^l|$ . Let

$$\tilde{K}_q^l \triangleq \min_{j\in G_q}\{K_j^l\} . \tag{23}$$

Similar to the analysis in previous subsection, by using the Hoeffding's tail inequality [38][39], we have

$$\Pr\left(\boldsymbol{\alpha}_q^l(n) > \boldsymbol{\alpha}_q^l(m)\right) \ge 1 - e^{-\frac{|G_q|\left(p_{1,q}^l - p_{2,q}^l\right)^2}{2}} , \tag{24}$$

for $n \in T$ and $m \in T_c$ , where

$$\begin{aligned} p_{1,q}^l &\triangleq \frac{\tilde{K}_q^l}{K} \\ p_{2,q}^l &\triangleq \frac{K-\tilde{K}_q^l}{N-K} \end{aligned} . \tag{25}$$

Note from (22) that

$$\tilde{K}_q^l > \tilde{K}_q^{l-1} . \tag{26}$$

From (15), (25) and (26), it follows

$$p_{1,q}^l > p_{1,q}^{l-1} > p_{2,q}^{l-1} > p_{2,q}^l . \tag{27}$$

Thus, it can be concluded that the lower bound of $\Pr\left(\boldsymbol{\alpha}_q^l(n) > \boldsymbol{\alpha}_q^l(m)\right)$ keeps increasing during the iterative process, i.e.,

$$\Pr\left(\boldsymbol{\alpha}_q^l(n) > \boldsymbol{\alpha}_q^l(m)\right)_{\text{lower bound}} = 1 - e^{-\frac{|G_q|\left(p_{1,q}^l - p_{2,q}^l\right)^2}{2}} > 1 - e^{-\frac{|G_q|\left(p_{1,q}^{l-1} - p_{2,q}^{l-1}\right)^2}{2}} = \Pr\left(\boldsymbol{\alpha}_q^{l-1}(n) > \boldsymbol{\alpha}_q^{l-1}(m)\right)_{\text{lower bound}} , \tag{28}$$

for $n \in T$ and $m \in T_c$ .

Both Proposition 1 and (28) imply that, as the iterations proceed, every element in $T_q^l$ is correct with high probability when the number of neighboring nodes is sufficiently large. Therefore,

$$\left\| \mathbf{x}_q(T - T_q^l) \right\|_2 \le \left\| \mathbf{x}_q(T - \Omega_q^l) \right\|_2 \tag{29}$$

holds with high probability when the value of $|G_q|$ satisfies (5). From (21) and (29), it follows that

$$\left\| \mathbf{x}_q(T - T_q^l) \right\|_2 < \left\| \mathbf{x}_q(T - T_q^{l-1}) \right\|_2 , \tag{30}$$

holds with high probability. This implies that DCSP with sufficient number of one-hop communication nodes ensures the continuous reduction of error of support set estimation. This completes the proof of the convergence of DCSP.

It is worth emphasizing that the actual number of neighbors of each node required for DCSP to successfully recover the joint sparsity pattern may be much smaller than the sufficient condition in (5), as shown by simulations in Section V. The reasons are: 1) the sufficient condition in (5) corresponds to the worst case scenario at the initialization phase, i.e., $\tilde{K}_q^0 = 1$; and 2) in the above theoretical analysis, it is assumed that $K_j^l = \tilde{K}_q^l = \min_{j \in G_q}\{K_j^l\}$ for all $j \in G_q$ at the $l$-th iteration. When $\tilde{K}_q^0 > 1$ and $K_j^l > \tilde{K}_q^l$ occur, i.e., when at some nodes the accuracy of local support set estimate is better than expected, the number of neighboring nodes required to guarantee the desired fusion performance will decrease.

### *B. Communication overhead of DCSP*

As described above, fusion of the local support set estimates in DCSP increases the number of correct indices selected at each iteration. In the following, we investigate the number of iterations needed for DCSP to successfully recover the support set.

From (6), (18), (21) and (29), it follows that

$$\left\| \mathbf{x}_q(T - T_q^l) \right\|_2 \le \left[ \frac{2\delta_{3K}(1+\delta_{3K})}{(1-\delta_{3K})^3} \right]^l \frac{\sqrt{8\delta_{3K} - 8\delta_{3K}^2}}{1+\delta_{3K}} \left\| \mathbf{x}_q \right\|_2 \tag{31}$$

holds with high probability, when the number of one-hop neighbors of the $q$-th node is sufficiently large. A necessary and sufficient condition for $T = T_q^l$ is that $\left\| \mathbf{x}_q(T - T_q^l) \right\|_2 < \min\{|\mathbf{x}_q(i)|, i \in T\}$. Hence, the number of iterations required for DCSP at the $q$-th node to accurately recover the joint sparsity pattern can be expressed by

$$L_{DCSP} = \left\lceil \frac{\log\left(\alpha \cdot \frac{1+\delta_{3K}}{\sqrt{8\delta_{3K} - 8\delta_{3K}^2}}\right)}{\log\left(\frac{2\delta_{3K}(1+\delta_{3K})}{(1-\delta_{3K})^3}\right)} \right\rceil, \tag{32}$$

where $\lceil \cdot \rceil$ is the ceiling function and

$$\alpha \triangleq \frac{\min\{|\mathbf{x}_q(i)|, i \in T\}}{\|\mathbf{x}_q\|_2}. \tag{33}$$

It is clear that the value of $L_{DCSP}$ depends on the decay of the nonzero coefficients of the sparse signal. Consider a special case of random coefficients of {$\mathbf{x}_q(T)$, $q$=1, 2, ⋯, $Q$}, where the coefficients of $\mathbf{x}_q(T)$ are generated from the same decay function but arranged in random order for $q$=1, 2, ⋯, $Q$. In this case, the value of $L_{DCSP}$ can be exactly determined by (32). For example, for the sparse signals with exponentially decaying entries, in which the $n$-th large amplitude is constrained by $c_{\exp} \cdot e^{-p_{\exp}(n-1)}$ for some constants $c_{\exp} > 0$ and $p_{\exp} > 0$, (33) can be specified as

$$\alpha_{\exp} = e^{-p_{\exp}(K-1)} \cdot \sqrt{\frac{1-e^{-2p_{\exp}}}{1-e^{-2p_{\exp}K}}}. \tag{34}$$

For sparse signals with power-law decaying entries, in which the $n$-th large amplitude is constrained by $c_{power} \cdot n^{-p_{power}}$ for some constants $c_{power} > 0$ and $p_{power} > 1$, (33) becomes

$$\alpha_{power} = \frac{K^{-p_{power}}}{\sqrt{\sum_{n=1}^{K} n^{-2p_{power}}}}. \tag{35}$$

Based on (32)-(35), in Fig. 2 we plot the number of iterations required for DCSP to exactly recover the joint sparsity pattern versus the varying RIP constant $\delta_{3K}$. The parameters of the exponentially decaying function are $c_{\exp} = 1$ and $p_{\exp} = 0.3$, while the parameters of the power-law decaying function are $c_{power} = 1$ and $p_{power} = 1.5$. For both kinds of sparse signals, the sparsity $K$ is fixed at 10. It is clear from Fig. 2 that the required number of iterations of DCSP is of the order of O($K$). Moreover, when the value of $\delta_{3K}$ goes down (approximately corresponding to the increase in the number of measurements per node), DCSP may recover the joint sparsity pattern via only one iteration.

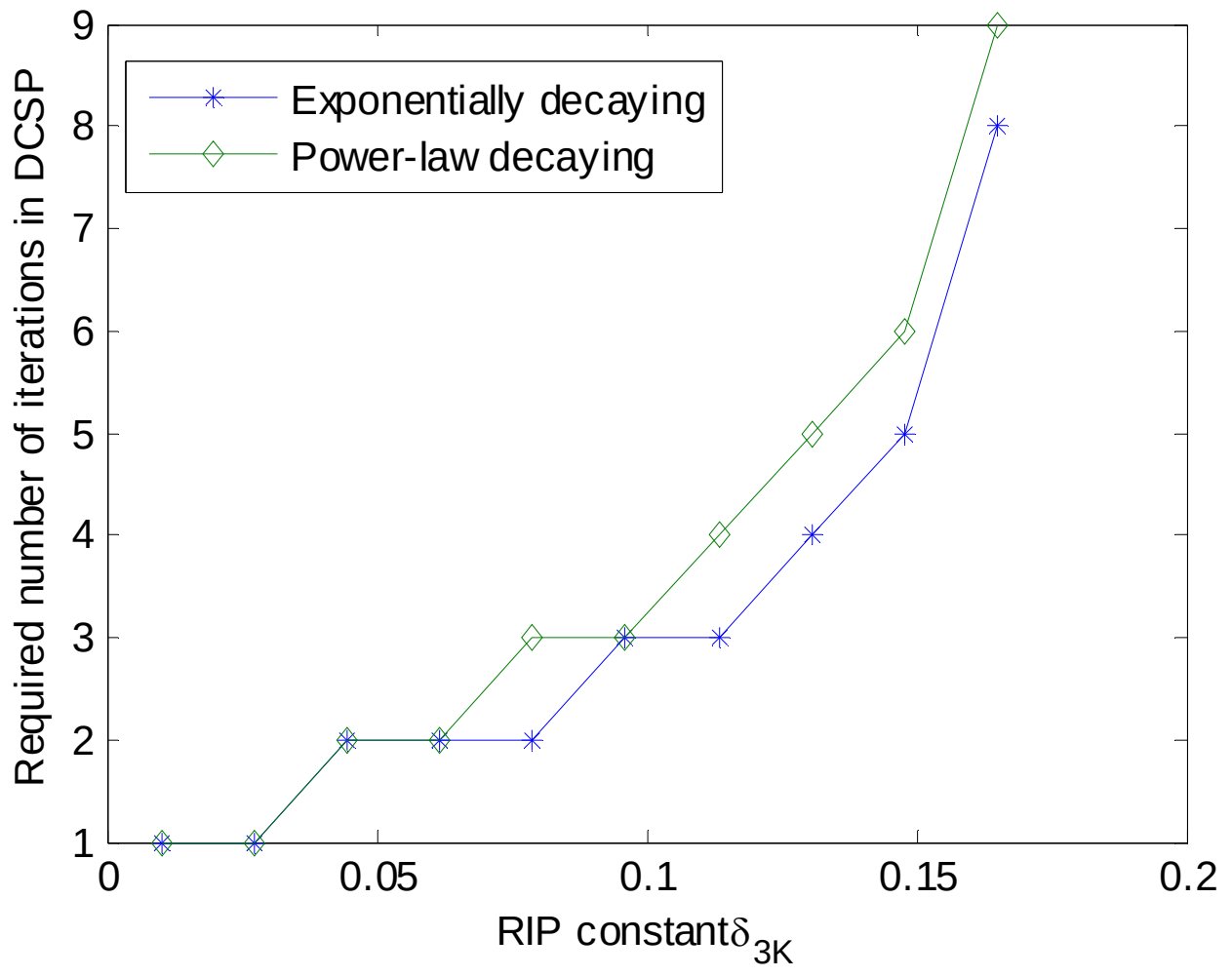

Fig. 2 The required number of iterations in DCSP versus RIP constant $\delta_{3K}$ . $K$=10.

With the bound on the number of iterations, the communication overhead of DCSP can be easily estimated. From Algorithm 1, we can see that the total number of messages to be transmitted in DCSP is

$$C_{DCSP} = (K + KL_{DCSP} + L_{DCSP})\sum_{q=1}^{Q}\left(\left|G_q\right| - 1\right), \tag{36}$$

where the bound of $L_{DCSP}$ is given in (32). The experimental results in Fig. 2 and Section V show that $L_{DCSP}$ is of the order of O($K$). Thus, if each node is assumed to communicate with its one-hop neighbors one by one at each iteration, the communication overhead of DCSP is about $O(K^2)\sum_{q=1}^{Q}\left(\left|G_q\right| - 1\right)$.

## IV. GENERALIZED DCSP ALGORITHM FOR JOINT SPARSITY PATTERN RECOVERY

In DCSP proposed in the previous section, the collaboration among nodes is based on the exchanges of $K$-length indices. For small-scale networks (e.g., cognitive radio networks with several nodes), it is reasonable to assume that one-hop communication between all pairs of nodes in the network is available. Under this assumption, we investigate how to further improve the accuracy of the estimation of the sparsity pattern. In [18], [20], [28] and [29], it was shown that the accuracy of the sparsity pattern recovery can be improved by allowing the transmission of O($N$)-length correlation coefficients and projection coefficients among some nodes. Based on this observation, we propose the following GDCSP algorithm, in which each node is allowed to share O($N$)-length messages and $K$-length indices with a few neighboring nodes and other nodes in the network, respectively. By doing so, the accuracy of support set estimation of GDCSP is expected to be improved in comparison with DCSP, at the expense of increased communication overhead. In GDCSP, we denote $G_q$ as the surrounding neighborhood of the $q$-th node in which the transmission of O($N$)-length vectors is allowed. The GDCSP algorithm is summarized in Algorithm 2. There are two kinds of communication in GDCSP, i.e., local communication and global communication. In Steps 1 and 4, the $q$-th node shares an $N$-length correlation coefficient vector $\mathbf{c}_q^l$ with its neighbors. The coefficients in $\mathbf{c}_q^l$ are probably not sparse before the iterations converge, so the transmission of the entire $\mathbf{c}_q^l$ is necessary. The neighboring nodes collaborate once again in Step 6 of GDCSP, where the $q$-th node shares a 2$K$-length projection coefficient vector $\mathbf{d}_q^l$ with its neighbors. By such two-stage collaboration, the $q$-th node obtains $\Omega_q^l$ as the local estimate of the support set by finding a $K$-dimensional subspace that the measurement vectors collected within the neighborhood most probably lies in. Then, all the nodes in the network share local estimates of the support set with each other to enable global fusion based on majority vote, as described in Steps 3 and 7 of GDCSP. When $G_q= G$ for $q$=1, 2, ⋯, $Q$, GDCSP is equivalent to the centralized SSP algorithm proposed in [20].

---

**Algorithm 2 The GDCSP algorithm at the *q*-th node**

Input: $K$, $\mathbf{y}_q$ , $\mathbf{A}_q$ .

Initialization:

1) Transmit the vector $\mathbf{c}_q^0 = |\mathbf{A}_q^H \mathbf{y}_q|$ to and receive $\mathbf{c}_j^0$ from the $j$-th node for all $j \in G_q \backslash \{q\}$.

2) Let $\Omega_q^0$ =max_ind( $\sum_{j\in G_q} \mathbf{c}_j^0$ , $K$). Send $\Omega_q^0$ to and receive $\Omega_j^0$ from the $j$-th node for all $j \in G \backslash \{q\}$.

3) Let $\Gamma_q^0 = \{\Omega_j^0, j \in G\}$ and $T_q^0$ =max_occ( $\Gamma_q^0$ , $K$). Calculate the local residual $\mathbf{r}_q^0$ = resid($\mathbf{y}_q$, $\mathbf{A}_q$( $T_q^0$ )).

Iteration: at the $l$-th iteration ($l \geqslant 1$)

4) Send the vector $\mathbf{c}_q^l = |\mathbf{A}_q^H \mathbf{r}_q^{l-1}|$ to and receive $\mathbf{c}_j^l$ from the $j$-th node, for all $j \in G_q \backslash \{q\}$.

5) Calculate the projection coefficients $\mathbf{d}_q^l = |\mathrm{ls}(\mathbf{y}_q, \mathbf{A}_q(\tilde{T}_q^l))|$, where $\tilde{T}_q^l = T_q^{l-1} \cup$ max_ind( $\sum_{j\in G_q} \mathbf{c}_j^l$ , $K$) .

6) Send $\mathbf{d}_q^l$ to and receive $\mathbf{d}_j^l$ from the $j$-th node for all $j \in G_q \backslash \{q\}$.

7) Let $\Omega_q^l$ =max_ind( $\sum_{j \in G_q} \mathbf{d}_j^l$ , $K$). Send $\Omega_q^l$ to and receive $\Omega_j^l$ from the $j$-th node for all $j \in G \backslash \{q\}$.

8) Let $\Gamma_q^l = \{\Omega_j^l, j \in G\}$ and $T_q^l$ =max_occ( $\Gamma_q^l$ , $K$). Update the residual $\mathbf{r}_q^l$ = resid($\mathbf{y}_q$, $\mathbf{A}_q$( $T_q^l$ )). Send $\left\|\mathbf{r}_q^l\right\|_2$ to and receive $\left\|\mathbf{r}_j^l\right\|_2$ from the $j$-th node for all $j \in G \backslash \{q\}$.

9) If $\sum_{q=1}^{Q} \left\|\mathbf{r}_q^l\right\|_2 \geq \sum_{q=1}^{Q} \left\|\mathbf{r}_q^{l-1}\right\|_2$ , let $T_q^l = T_q^{l-1}$ and stop the iteration; otherwise, let $l$=$l$+1, and return to Step 4.

Output: The estimated support set $\hat{T}_q = T_q^l$ .

---

The required number of iterations for GDCSP to exactly recover the support set is expected to be slightly smaller than that in DCSP, since sharing of O($N$)-length coefficients among neighboring nodes improves the accuracy of local estimation of the support set. Let $L_{GDCSP}$ denote the required number of iterations for GDCSP. Then the communication cost of GDCSP can be estimated to be

$$C_{GDCSP} = N \sum_{q=1}^{Q} \left( \left| G_q \right| - 1 \right) + KQ(Q-1) + L_{GDCSP} \left( (N+2K) \sum_{q=1}^{Q} \left( \left| G_q \right| - 1 \right) + (K+1)Q(Q-1) \right). \quad (37)$$

The experimental results in Section V show that $L_{GDCSP}$ is of the order of O($K$), so the communication overhead of GDCSP is about $O(KN) \sum_{q=1}^{Q} \left( \left| G_q \right| - 1 \right) + O(K^2) Q(Q-1)$ .

In what follows, we summarize the communication overheads of some distributed greedy algorithms including DCSP and GDCSP. As shown in [27] and [28], the number of iterations of DCOMP, DiOMP and DiSP are also of the order of O($K$). At each iteration of DCOMP, each node transmits its $N$-length coefficients to surrounding neighbors for local communication and one index to all other nodes for global communication, respectively [28]. DiOMP and DiSP require each node to send $K$ indices to its one-hop neighbors per iteration [27]. The communication overheads of different algorithms are summarized in Table 1, which is consistent with the simulations results presented in Section V.

Table 1 Communication overhead of different algorithms.

| Algorithm | Communication overhead |
|---|---|
| DCSP | $O(K^2)\sum_{q=1}^{Q}\left(\lvert G_q\rvert-1\right)$ |
| GDCSP | $O(KN)\sum_{q=1}^{Q}\left(\lvert G_q\rvert-1\right)+O(K^2)Q(Q-1)$ |
| DCOMP | $O(KN)\sum_{q=1}^{Q}\left(\lvert G_q\rvert-1\right)+O(K)Q(Q-1)$ |
| DiOMP | $O(K^2)\sum_{q=1}^{Q}\left(\lvert G_q\rvert-1\right)$ |
| DiSP | $O(K^2)\sum_{q=1}^{Q}\left(\lvert G_q\rvert-1\right)$ |

## V. EXPERIMENTAL RESULTS

In this section, simulation results are provided to demonstrate the performance of the proposed DCSP and GDCSP algorithms. Without loss of generality, two assumptions are made on the network configuration: 1) $|G_q|=g<Q$ for all the nodes; 2) the neighbors of the $q$-th node are indexed by {mod($q$+1,$Q$), mod($q$+2,$Q$),⋯, mod($q$+$g$-1,$Q$)}, where mod(·) is the modulus operation.

### *A. When the sparse signals at all the nodes have the same sparsity pattern but different coefficients*

We first evaluate the performance of DCSP and GDCSP in terms of the accuracy of sparsity pattern recovery. Signal sparsity is fixed at $K$=10, the length of the sparse signal at each node is set to $N$=200, and the number of nodes is $Q$=10. We randomly generate a set of $M\times N$ dictionary matrices $\{\mathbf{A}_q, q=1, 2, \cdots, Q\}$ from the standard independent and identically distributed (iid) Gaussian ensemble. Next, we randomly select $K$ indices from $\{1, 2,\cdots, N\}$ as the support set $T$ and draw the entries of $\mathbf{x}_q$ supported on $T$ from the standard iid Gaussian ensemble. Then, we generate the measurement vectors $\mathbf{y}_q=\mathbf{A}_q\mathbf{x}_q+\mathbf{w}_q$ for $q$=1, 2,⋯, $Q$, where $\mathbf{w}_q$ is the additive Gaussian noise. The noise is assumed to be statistically independent of the signals. The signal-to-noise ratio (SNR) is defined as $\text{SNR} \triangleq \sum_{q=1}^{Q}\left\|\mathbf{x}_q\right\|_2^2/(QN\sigma_{\mathrm{w}}^2)$, where $\sigma_{\mathrm{w}}^2$ is the variance of the noise. With the same parameter setting, we apply different algorithms to estimate the common support set. If the estimation result $\hat{T}_q = T$ for all the nodes, sparsity pattern recovery is considered as successful. In Figs. 3 and 4, we let SNR= 18dB and evaluate the success rates of different algorithms versus the number of measurements per node by averaging over 500 Monte Carlo trials. Fig. 3 provides the trends of accuracies of the proposed DCSP and GDCSP algorithms as the neighborhood size $g$ increases. The performance of the centralized SSP algorithm [20], which is equivalent to GDCSP with $g=Q$, is also given as a benchmark. As expected, the accuracies of DCSP and GDCSP increase as each node collaborates with more neighbors. Moreover, the more neighbors each node can collaborate with, the less measurements per node are required. When $g$ gets closer to $Q$, the accuracy of DCSP approaches that of the centralized SSP algorithm. In Fig. 4, we set the neighborhood size $g$=5 and compare DCSP and GDCSP with three other decentralized algorithms , DCOMP [28][1], DiOMP, and DiSP [27], in terms of accuracy of sparsity pattern recovery. The benchmark performances of the centralized SOMP [18] and SSP [20] algorithms are given as well. If the bandwidth of the network is extremely limited so that the transmission of $N$-length messages is not permitted, only DiOMP, DiSP and DCSP are suitable for this scenario. Unlike DiOMP that deems all the indices selected during past iterations reliable, at each iteration DCSP is capable of removing wrong indices selected earlier and adding new index candidates into the support set estimate. Unlike DiSP that triggers the index exchange among neighboring nodes after the local sparse solution is fully recovered at every node, DCSP activates the collaboration among neighboring nodes before the local solution process converges. By doing so, the local recovery errors at each node and, therefore, the performance loss of fusion among neighboring nodes are mitigated. Therefore, as shown in Fig. 4, DCSP algorithm offers much better accuracy of sparsity pattern recovery compared to DiOMP and DiSP. In the case where each node can communicate with all others in the network via one-hop communication, both DCOMP and GDCSP require the exchange of $N$-length coefficients among neighboring nodes and the exchange of O($K$)-length indices among remote nodes. From Fig. 4, we can see that the accuracy of GDCSP is much better than that of DCOMP and very close to that of the centralized SSP algorithm. In particular, Fig. 4 also indicates that, DCSP, which is based on the local transmission of $K$-length messages, is more accurate than DCOMP, which is based on the local transmission of $N$-length messages and requires global connectivity of the network.

---

[1] Here the term DCOMP refers to the DC-OMP 2 algorithm that has better accuracy than DC-OMP 1 [28].

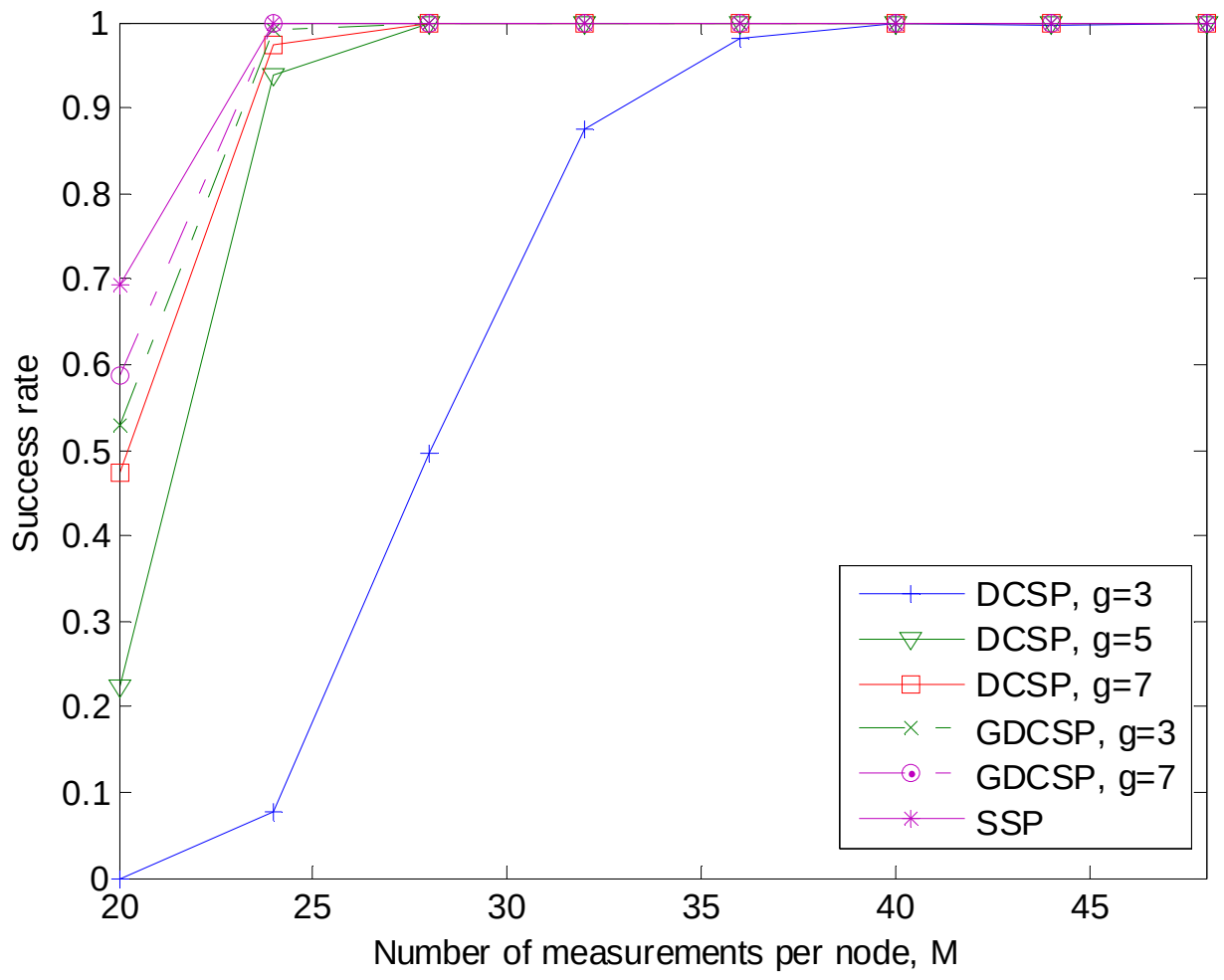

Fig. 3 Performance of DCSP and GDCSP with increasing neighborhood size, for independent sparse signal model. $Q$=10, $K$=10, $N$=200, and SNR = 18dB.

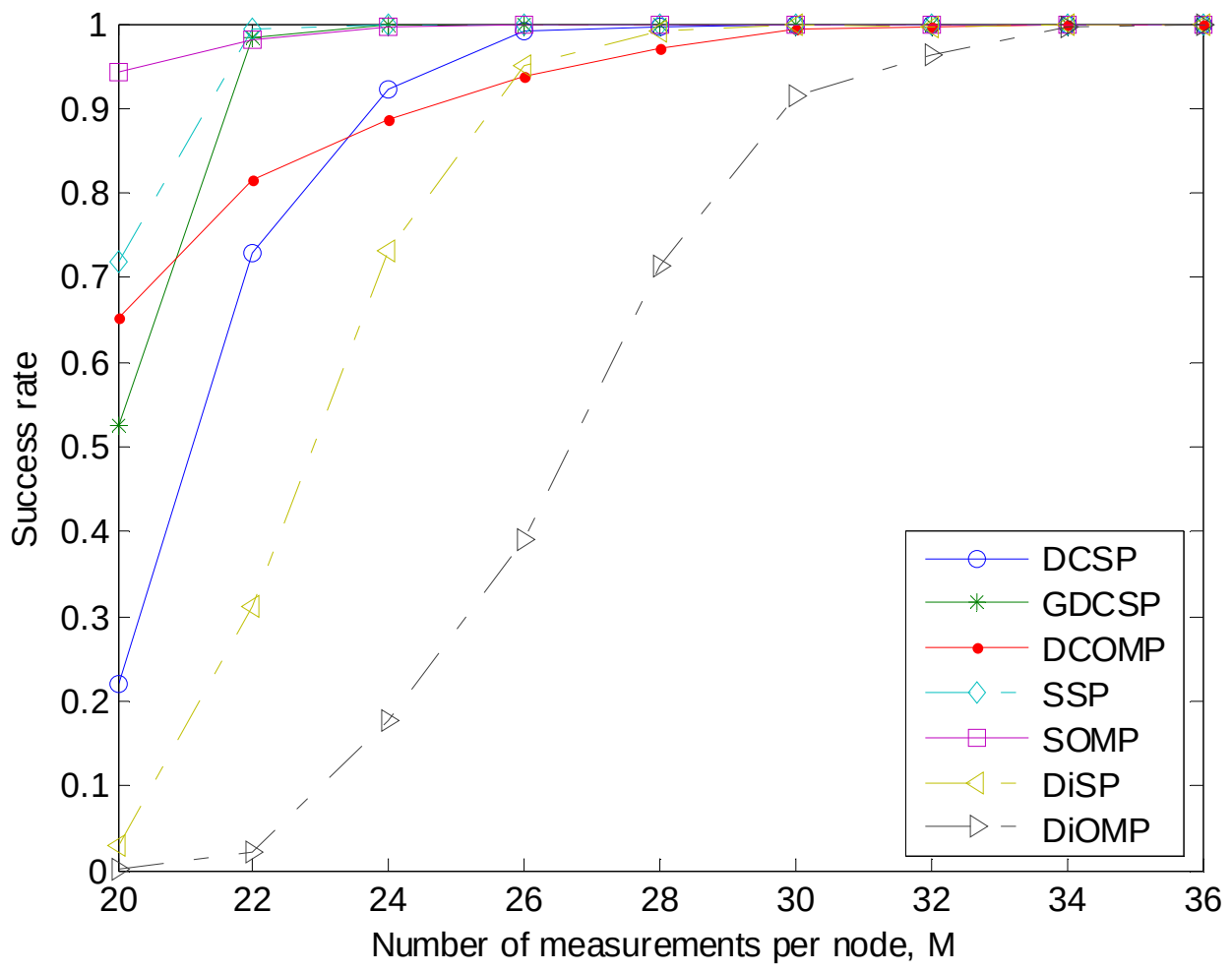

Fig. 4 The success rates of multiple algorithms versus the number of measurements per node, for independent sparse signal model. $Q$=10, $K$=10, $N$=200, $g$=5 and SNR = 18dB.

In the next experiment, the comparison among five decentralized algorithms, DCOMP, DiOMP, DiSP, DCSP, and GDCSP, in terms of communication cost is shown. Let SNR=18 dB, $M$=40, $K$=10, and $N$=200. Assume that the network size is increasing, i.e., $Q$ is varying from 10 to 100, and the neighborhood size $g$ equals $Q$/2 for all the nodes. In Figs. 5 and 6, the average number of iterations and the communication overheads required for different algorithms to exactly recover the sparsity pattern are plotted by averaging over 100 Monte Carlo trials, respectively. Thanks to the efficient selection of indices at each node and the effective collaboration among nodes, DiSP, DCSP and GDCSP have faster convergence speeds, as shown in Fig. 5. From Fig. 6 we can see: 1) among the algorithms that only require exchange of *K*-length messages among nodes, the communication overhead of DCSP is much less than that of DiOMP and comparable with that of DiSP; and 2) if the transmission of *N*-length messages among neighboring nodes is allowed, the communication efficiency of GDCSP is slightly better than that of DCOMP. In particular, DCSP, DiSP and GDCSP can even recover the joint sparsity pattern from only one iteration when the number of neighbors of each node is sufficiently large. This means that, fusion in the initialization phase of DCSP has correctly recovered the support set, and only one iteration is needed to check whether the algorithm should be terminated. Both Figs. 4 and 6 indicate that, compared to existing decentralized greedy algorithms, DCSP and GDCSP improve the accuracy of sparsity pattern recovery while incurring comparable communication cost.

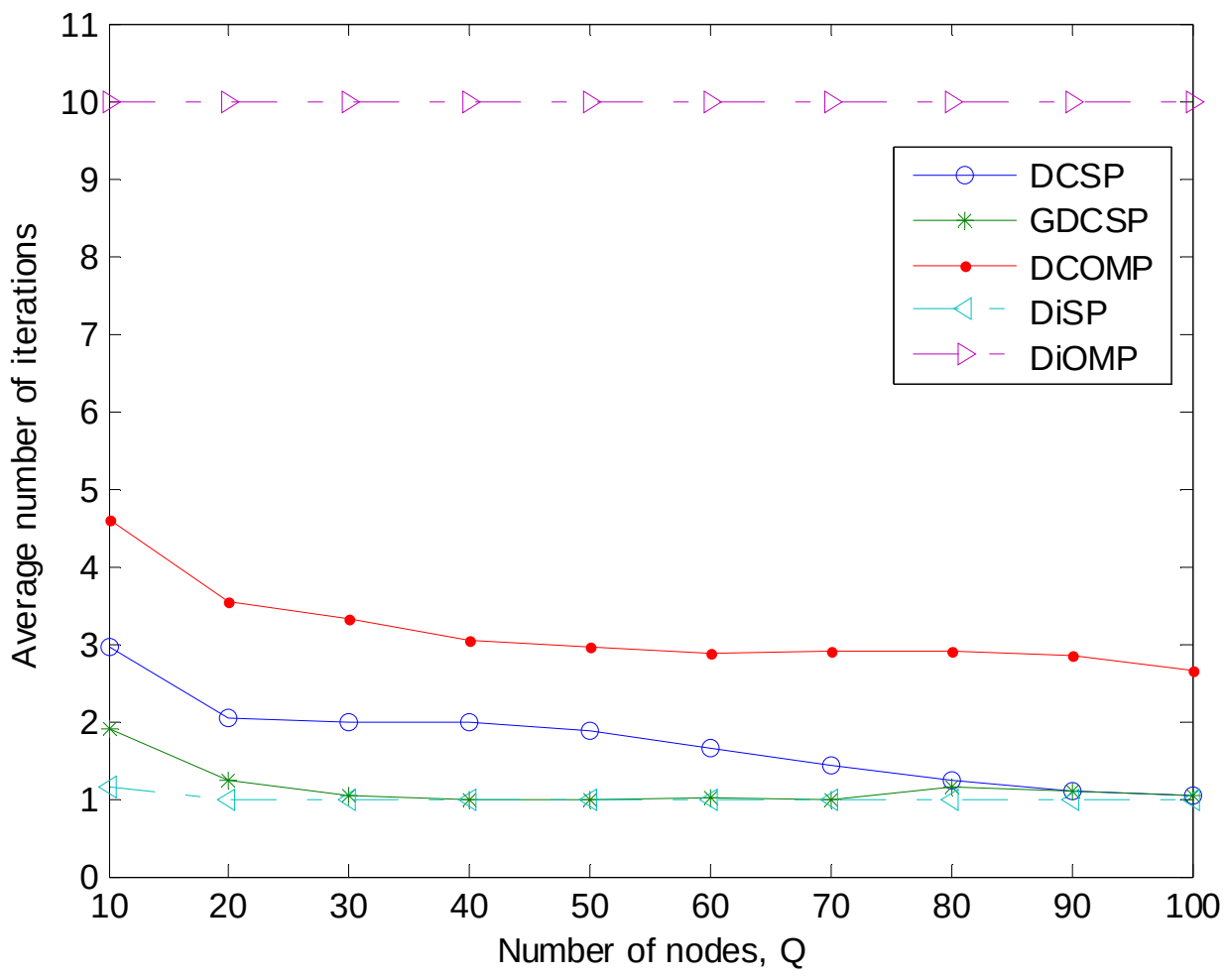

Fig. 5 The number of required iterations versus network size, for independent sparse signal model. *M*=40, *N*=200, *K*=10 and SNR=18dB.

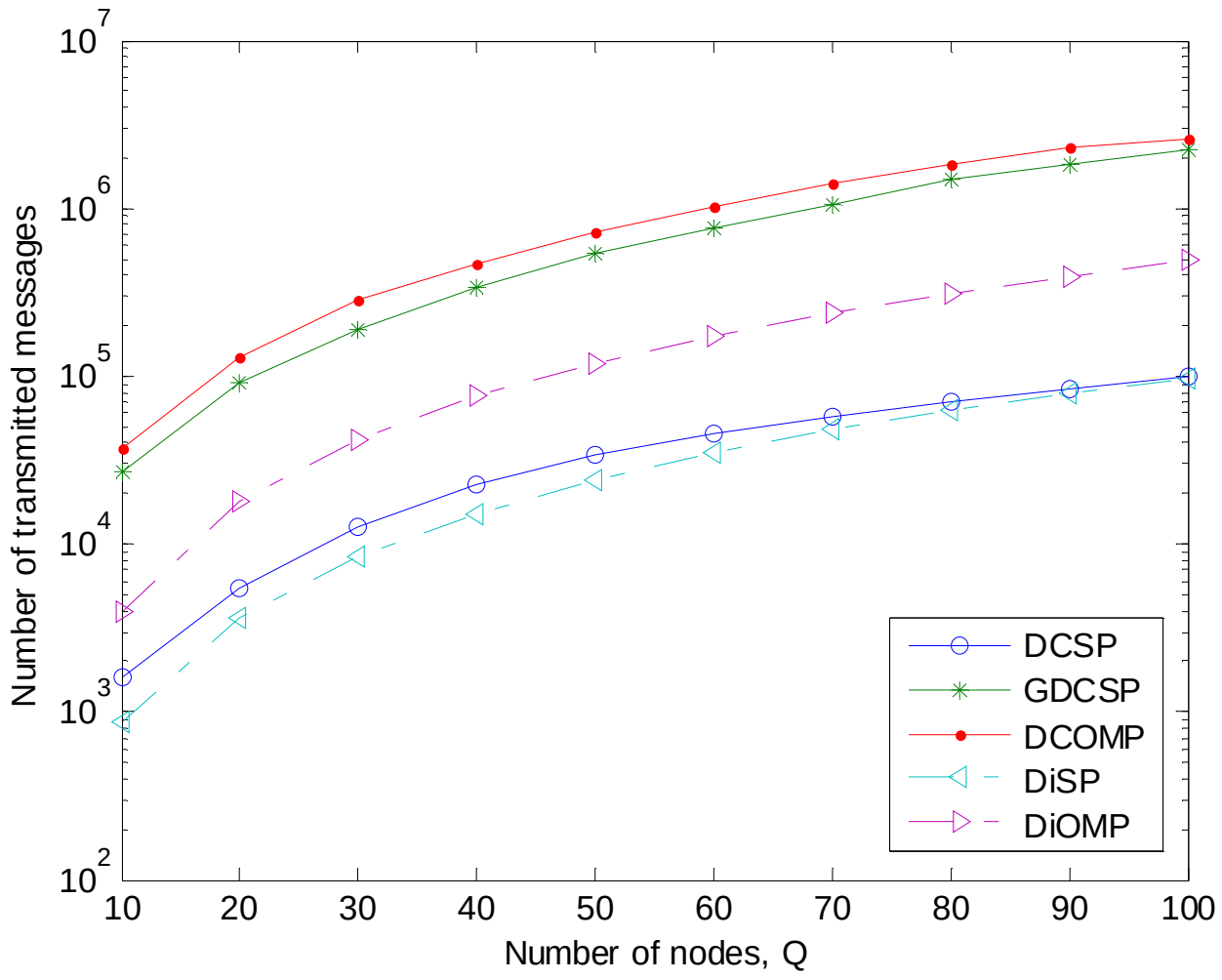

Fig. 6 The number of messages to be transmitted versus network size scale, for independent sparse signal model. $M$=40, $N$=200, $K$=10 and SNR=18dB.

### *B. When all the nodes observe the same sparse signal*

Although the analysis in Section III was performed for the scenario where the sparse signals at different nodes were statistically independent, we would like to experimentally evaluate the performance of the proposed DCSP and GDCSP algorithms when the sparse signals at all the nodes are the same. This signal model can be expressed as $\mathbf{y}_q=\mathbf{A}_q\mathbf{x}+\mathbf{w}_q$, in which $K$ nonzero coefficients of the sparse signal $\mathbf{x}$ are supported on $T$. Let $N$=200, $K$=10, $Q$=10, $g$=5 and SNR = 18dB. In Fig. 7, we compare different algorithms in terms of the accuracy of sparsity pattern recovery with varying number of measurements. Simulations are carried out in a similar way as the first experiment, but here in each trial we randomly generate a $K$-sparse signal $\mathbf{x}$ and assume it to be available at all the nodes, i.e., $\mathbf{x}_q=\mathbf{x}$ for $q$=1, 2,⋯, $Q$. Compared to Fig. 4, the success rate of every algorithm in Fig. 7 is slightly increased. That is to say, when all the nodes observe the same sparse signal, these algorithms offer better accuracy of sparsity pattern recovery than that in the case of statistically independent sparse signals. From Fig. 7, we also see the superiority of DCSP over DiOMP and DiSP in terms of the success rate, similar to what we observed in Fig. 4. One may note that distributed LASSO can also be used to solve the problem of sparsity pattern recovery when $\mathbf{x}_q=\mathbf{x}$ for $q$=1, 2,⋯, $Q$. Here, the performances of LASSO and other optimization-based algorithms are not provided due to their huge computational complexity over greedy algorithms. In addition, it was reported in [36] that, for sparsity pattern recovery with large Gaussian measurement matrices in high SNR scenarios, LASSO and OMP have almost identical accuracy.

The convergence speeds and the communication overheads of different algorithms with varying network size are plotted in Figs. 8 and 9, respectively. The neighborhood size is set as $|G_q|=g=Q/2$ for $q=1, 2,\cdots, Q$. Compared to Figs. 5 and 6 where the sparse signals at different nodes are independent, the number of iterations and the communication costs of these algorithms are slightly increased, except for DiOMP. With the same sparse signal at all the nodes, $\{\boldsymbol{\beta}_q^l, q=1, 2, \cdots, Q\}$ are probably similar to each other. This definitely improves the performance of fusion based on majority vote and, therefore, the accuracy of sparsity pattern recovery, as indicated in Fig. 7; however, this reduces the chance to select more than $\tilde{K}_q^l$ correct indices resulting from majority voting fusion at each iteration. This is the reason why most algorithms require more iterations and, therefore, larger communication overheads when all the nodes observe the same sparse signal, compared to that in the independent sparse signal case.

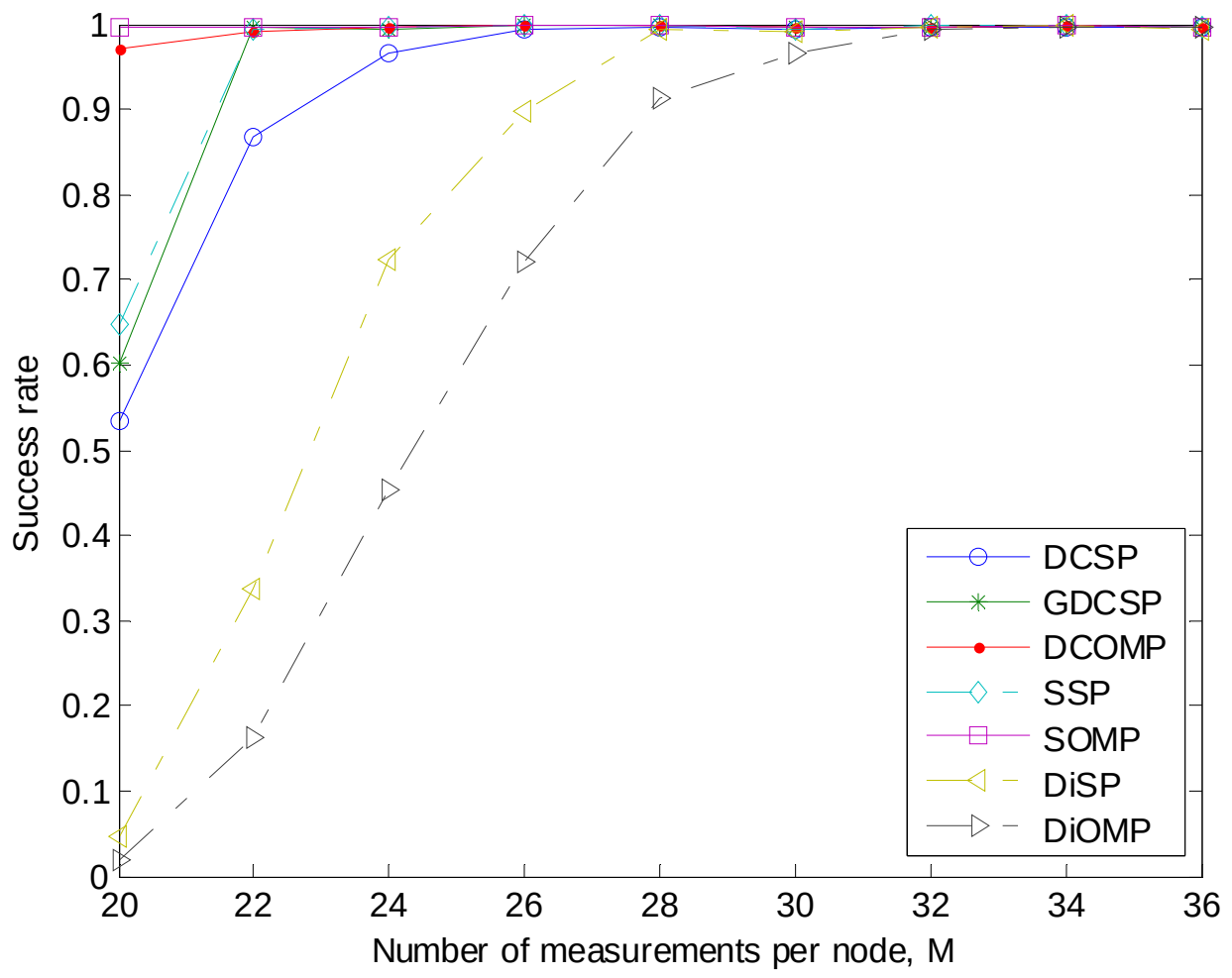

Fig. 7 The success rate of sparsity pattern recovery versus the number of measurements per node, when all the nodes observe the same sparse signal. $Q$=10, $K$=10, $N$=200, $g$=5 and SNR = 18dB.

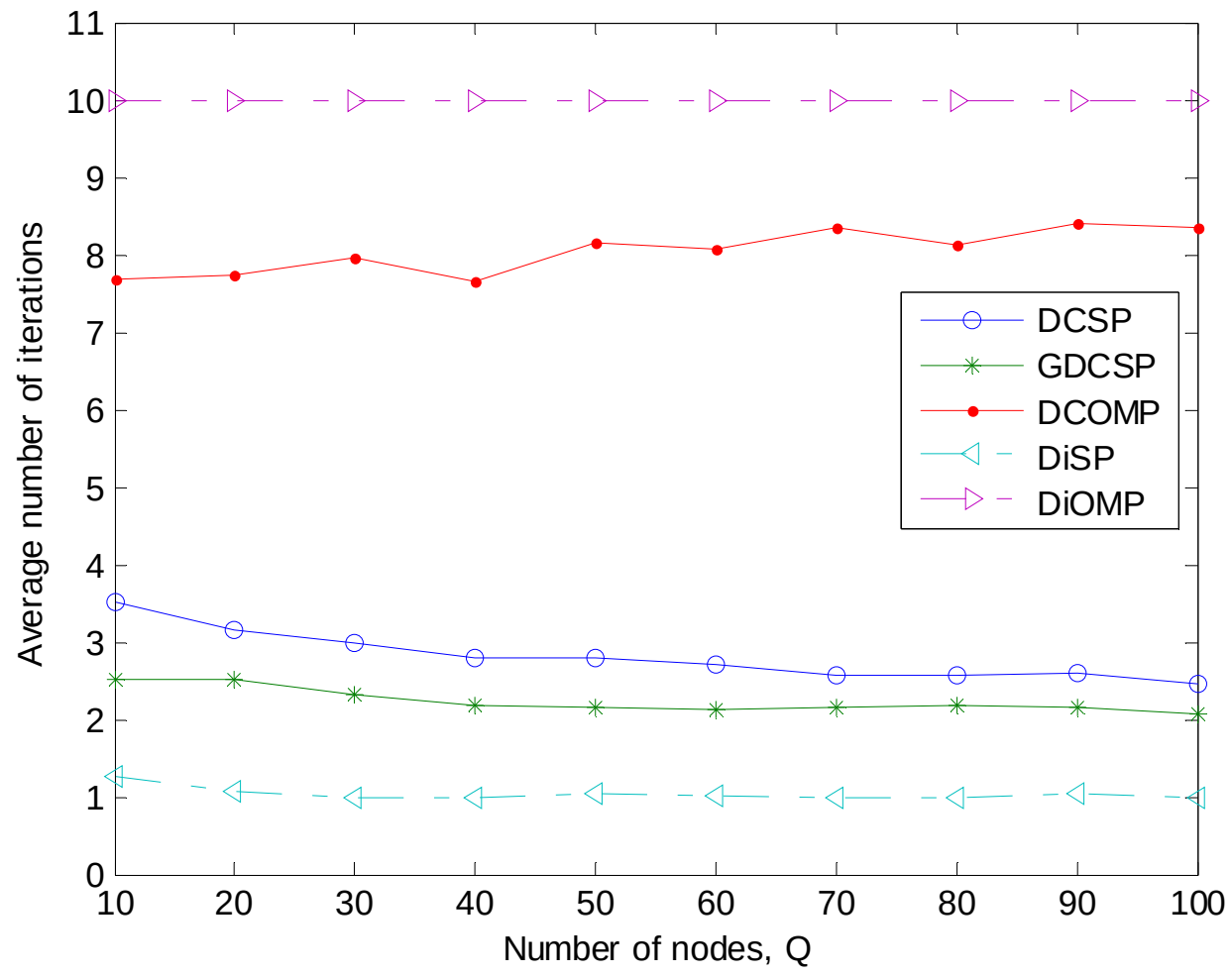

Fig. 8 The number of required iterations versus network size, when all the nodes observe the same sparse signal. $M$=40, $N$=200, $K$=10, and SNR = 18dB.

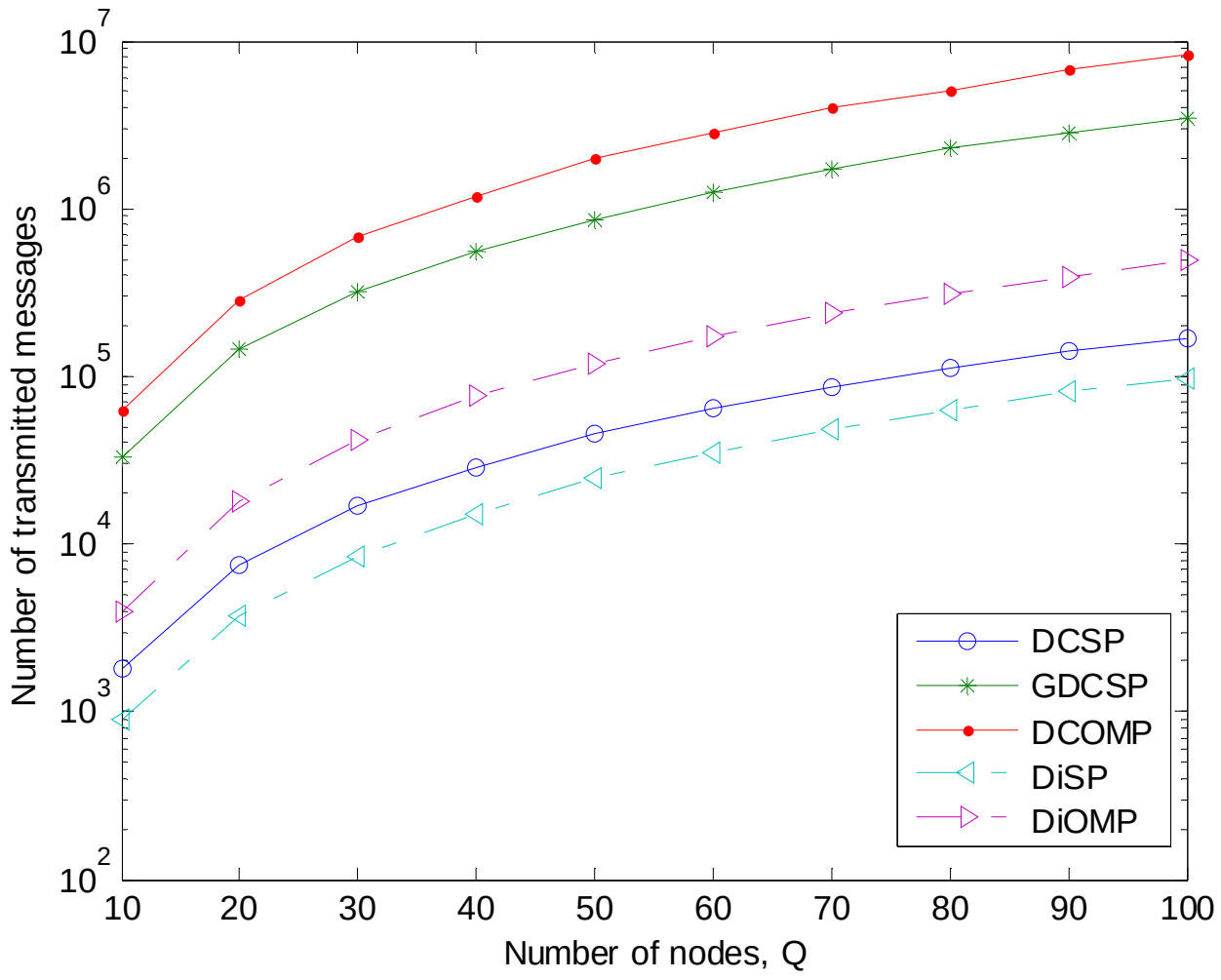

Fig. 9 The number of messages to be transmitted versus network size, when all the nodes observe the same sparse signal. $M$=40, $N$=200, $K$=10, and SNR = 18dB.

## VI. Conclusion

In this paper, we developed two decentralized greedy algorithms named DCSP and GDCSP for joint sparsity pattern recovery with a distributed sensor network without depending on a central fusion center. By embedding the exchanges of $K$-length support set estimates among one-hop neighboring nodes into the iterative operation of the standard SP algorithm, DCSP can accurately recover the common sparsity pattern with a small number of measurements per node as well as less communication cost of the network. In the scenario where one-hop communication between each pair of nodes in the network is available, GDCSP offers comparable accuracy as that of the centralized algorithm by allowing each node to share *N*-length coefficients with some of its surrounding neighbors. Experimental results show the superiority of DCSP and GDCSP over the existing decentralized greedy algorithms in terms of accuracy of sparsity pattern recovery with comparable computational loads. Our approach presented here can be easily combined with the compressive sampling matching pursuit (CoSaMP) algorithm [37], since CoSaMP and SP are very similar to each other. Our future work will include the extension of DCSP and GDCSP for structured sparse signals and specific network configurations.

## ACKNOWLEDGMENT

The authors are grateful to the reviewers for their thorough and insightful comments and suggestions.

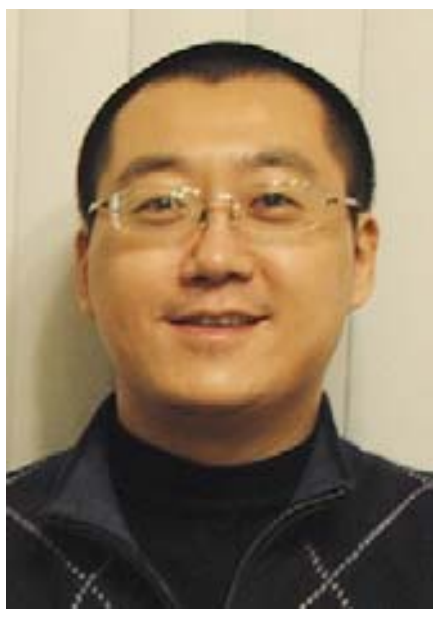

**Gang Li** (M’08–SM’13) received the B.S. and Ph.D. degrees in electronic engineering from Tsinghua University, Beijing, China, in 2002 and 2007, respectively. Since July 2007, he has been with the Faculty of Tsinghua University, where he is currently an Associate Professor with the Department of Electronic Engineering. From 2012 to 2014, he visited Ohio State University, Columbus, OH, USA, and Syracuse University, Syracuse, NY, USA. His research interests include radar imaging, time-frequency analysis, and compressed sensing.

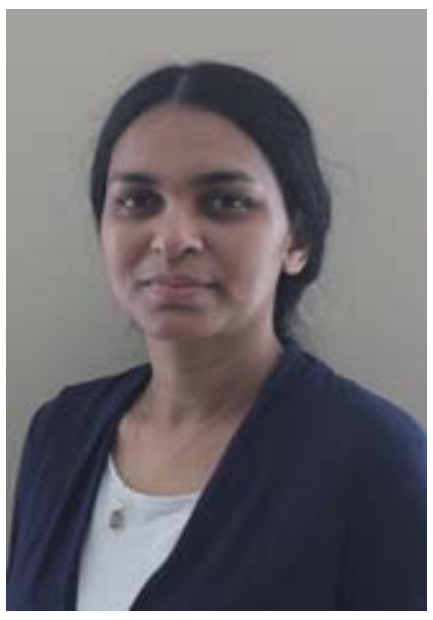

**Thakshila Wimalajeewa** (S’07–M’10) received the B.Sc. degree in electronic and telecommunication engineering (first class Hons.) from the University of Moratuwa, Moratuwa, Sri Lanka, in 2004, and the M.S. and Ph.D. degrees in electrical and computer engineering from the University of New Mexico, Albuquerque, NM, USA, in 2007 and 2009, respectively. From 2010 to 2012, she was a Postdoctoral Research Associate with the Department of Electrical Engineering and Computer Science, Syracuse University (SU), Syracuse, NY, USA. She currently holds a Research Faculty position at SU. Her research interests include the communication theory, signal processing, information theory, compressive sensing, low dimensional signal processing for communication systems, and resource optimization and decentralized processing in sensor networks.

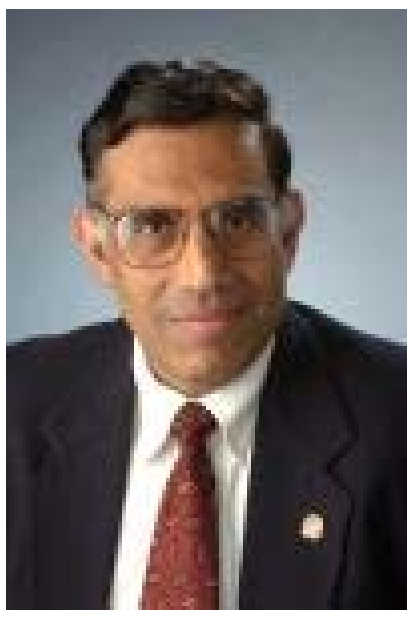

**Pramod K. Varshney** (S'72–M'77–SM'82–F'97) was born in Allahabad, India, on July 1, 1952. He received the B.S. degree in electrical engineering and computer science (with highest honors), and the M.S. and Ph.D. degrees in electrical engineering from the University of Illinois at Urbana-Champaign in 1972, 1974, and 1976 respectively. From 1972 to 1976, he held teaching and research assistantships with the University of Illinois. Since 1976, he has been with Syracuse University, Syracuse, NY, where he is currently a Distinguished Professor of Electrical Engineering and Computer Science and the Director of CASE: Center for Advanced Systems and Engineering. He served as the Associate Chair of the department from 1993 to 1996. He is also an Adjunct Professor of Radiology at Upstate Medical University, Syracuse. His current research interests are in distributed sensor networks and data fusion, detection and estimation theory, wireless communications, image processing, radar signal processing, and remote sensing. He has published extensively. He is the author of Distributed Detection and Data Fusion (New York: Springer-Verlag, 1997). Dr. Varshney was a James Scholar, a Bronze Tablet Senior, and a Fellow while at the University of Illinois. He is a member of Tau Beta Pi and is the recipient of the 1981 ASEE Dow Outstanding Young Faculty Award. He was elected to the grade of Fellow of the IEEE in 1997 for his contributions in the area of distributed detection and data fusion. He was the Guest Editor of the Special Issue on Data Fusion of the IEEE PROCEEDINGS January 1997. In 2000, he received the Third Millennium Medal from the IEEE and Chancellor's Citation for exceptional academic achievement at Syracuse University. He is the recipient of the IEEE 2012 Judith A. Resnik Award and Doctor of Engineering degree *honoris causa* from Drexel University in 2014. He is currently on the Editorial Boards of the Journal on Advances in Information Fusion and IEEE Signal Processing Magazine. He was the President of International Society of Information Fusion during 2001.